\documentclass[%
reprint,
groupedaddress,
 amsmath,amssymb,
 aps,
prl,
]{revtex4-1}
\usepackage{graphicx}% Include figure files
\usepackage{dcolumn}% Align table columns on decimal point
\usepackage{bm}% bold math
\usepackage[caption=false]{subfig}

\usepackage[colorinlistoftodos]{todonotes}

\begin{document}

\preprint{APS/123-QED}
\title{Transport in  Floquet-Bloch bands}
\author{C. J. Fujiwara}
\thanks{Equal contribution.}
\author{Kevin Singh\textsuperscript{*}}
\author{Zachary Geiger}
\author{Ruwan Senaratne}
\author{Shankari V. Rajagopal}
\author{Mikhail Lipatov}
\author{David M. Weld}
\email{weld@ucsb.edu}
\affiliation{University of California and California Institute for Quantum Emulation, Santa Barbara CA 93105, USA}
\begin{abstract}
We report Floquet band engineering of long-range transport  and direct imaging of Floquet-Bloch bands in an amplitude-modulated optical lattice. 
In one variety of Floquet-Bloch band we observe tunable rapid long-range high-fidelity transport of a Bose condensate across thousands of lattice sites. Quenching into an opposite-parity Floquet-hybridized band allows Wannier-Stark localization to be controllably turned on and off using modulation. A central result of this work is the use of transport dynamics to demonstrate direct imaging of a Floquet-Bloch band structure. These results demonstrate that transport in dynamical Floquet-Bloch bands can be mapped to transport in quasi-static effective bands, opening a path to cold atom quantum emulation of ultrafast multi-band electronic dynamics. %, new forms of Floquet engineering, and Floquet-enhanced metrology. 
\end{abstract}
\maketitle

Quantum control of transport in driven lattices may hold the key to new device types, unexplored techniques for ultrafast transport of energy and information in solids, and dynamical tools for controlling and probing condensed matter~\cite{floquetmatter,Holthaus2016,Eckardt2017,gedki-floquetbloch,Ghimire-HHGsolids,sherwin-recollision,ivanov-HHGbandstructure,huber-HHGblochoscs,Huber-realtimehhginsolids,reis-solidHHG,corkum-opticalbandreconstruction}. 
%Floquet band engineering is emerging as a powerful method for nonequilibrium material control and synthesis~\cite{floquetmatter,Holthaus2016,Eckardt2017}. 
%Cold atoms in optical lattices offer a near-ideal experimental platform for the study of nonequilibrium quantum transport~\cite{dahan1996,Haller2010,Geiger2018,Roati2008, Billy2008, Kondov66,sengstock-photoconductivity,Miller2007,Desbuquois2012,Stadler2012,Lin2011,Pedersen2013,Cheiney2013,Flaschner2018,Lignier2007,nagerlcorrelatedtunneling,weitz-tuningmobility,Parker2013,Alberti2009}.  
Ultracold atomic gases have enabled investigation of a wide variety of transport-related phenomena, including Bloch oscillations~\cite{dahan1996,
Haller2010,Geiger2018}, Anderson localization~\cite{Roati2008, Billy2008, Kondov66}, photoconductivity~\cite{sengstock-photoconductivity},  superfluid critical velocity~\cite{Miller2007,Desbuquois2012,Stadler2012}, spin-orbit coupling~\cite{Lin2011}, and pulsed modulation techniques for wavepacket manipulation~\cite{Pedersen2013,Cheiney2013}. The application of Floquet techniques in optical lattices~\cite{Holthaus2016,Eckardt2017} has expanded the control over these systems and enabled study of phenomena including topological dynamics~\cite{Flaschner2018}, renormalization of tunneling~\cite{Lignier2007}, correlated tunneling~\cite{nagerlcorrelatedtunneling}, tunable mobility~\cite{weitz-tuningmobility}, and  synthetic ferromagnets~\cite{Parker2013}.  While modulation techniques have been used to modify the spatial width of wavepackets~\cite{Alberti2009},  real-space control of center-of-mass transport  in Floquet-Bloch bands remains largely unexplored.

\begin{figure}[th!]
\centering
\includegraphics[width=0.9\linewidth]{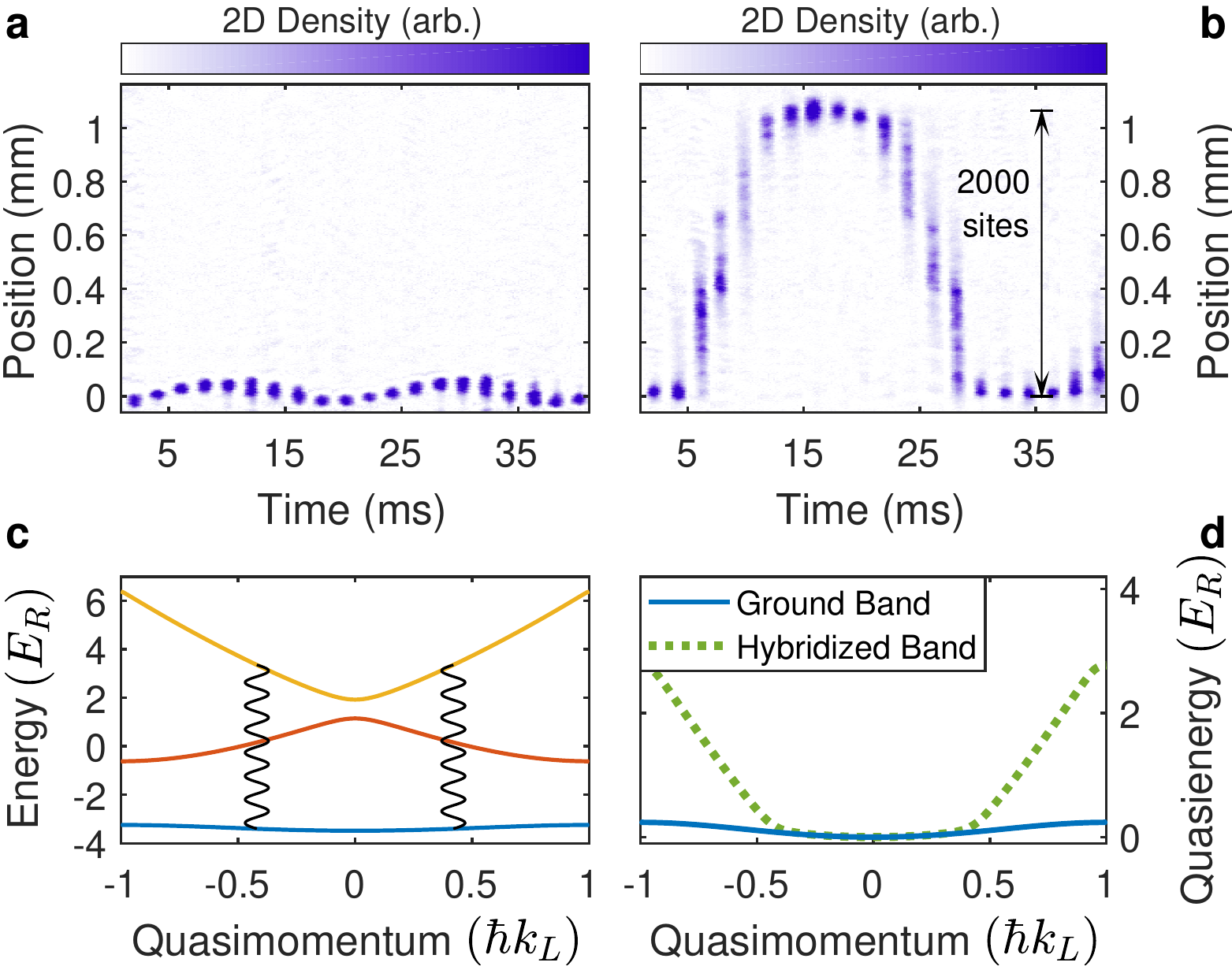}
\includegraphics[width=0.45\linewidth]{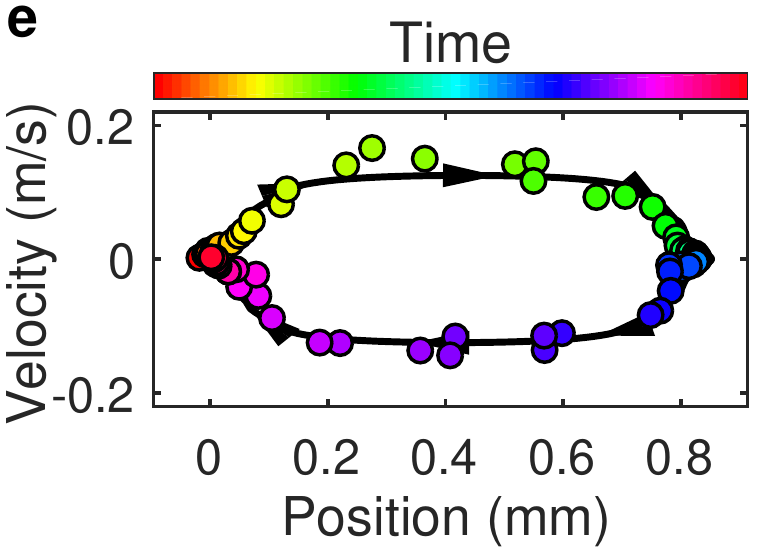}
\includegraphics[width=0.45\linewidth]{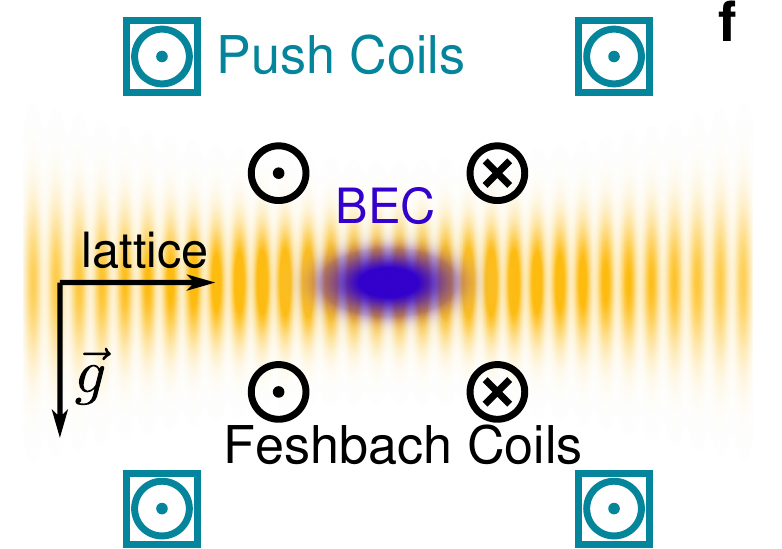}
\caption{Rapid long-range transport in a Floquet-Bloch band.  {\bf {(a)}} Time sequence of images of a condensate in the ground band of a $5.4~E_R$-deep lattice undergoing 48~Hz Bloch oscillations~\cite{Geiger2018}. {\bf (b)} Time sequence of images of a condensate in a $(0,2)$ hybridized Floquet-Bloch band created via amplitude modulation with $\nu=170$~kHz and $\alpha=0.2$, with the same initial force as in (a).  Note the rapid cyclic high-fidelity transport across the trap. {\bf (c)} Unmodified band structure. Vertical rippled lines indicate band coupling at the hybridizing quasimomentum for this modulation frequency. {\bf (d)} Calculated dispersion of the unmodified ground band (solid) and hybridized Floquet-Bloch band (dashed). {\bf (e)} Position-velocity evolution in the same hybrid band as (b) with an initial force corresponding to a Bloch frequency of 38~Hz. Solid line is theory; points are data at equally-spaced times. Error bars are smaller than plotted points.  {\bf (f)} Experimental schematic.}
\label{fig:fig1}
\end{figure}

%\caption{Rapid long-range transport in a Floquet-Bloch band.  {\bf {(a)}} Time sequence of images of a condensate in the ground band with lattice depth $V_0=5.4~E_R$. A force of 6$\times10^{-26}\,$N per atom induces Bloch oscillations with frequency 48~Hz~\cite{Geiger2018}. {\bf (b)} Time sequence of images of a condensate in a $(0,2)$ hybridized Floquet-Bloch band created via amplitude modulation with $\nu=170$~kHz and $\alpha=0.2$, with the same initial force as in (a).  Note the rapid cyclic high-fidelity transport across the trap. {\bf (c)} Unmodified band structure. Vertical rippled lines indicate band coupling at the hybridizing quasimomentum for this modulation frequency. {\bf (d)} Calculated dispersion of the unmodified ground band (solid) and hybridized Floquet-Bloch band (dashed). {\bf (e)} Position-velocity evolution in the same hybrid band as (b) with an initial force of 5$\times10^{-26}\,$N per atom corresponding to a Bloch frequency of 38~Hz. Solid line is theory; points are data, taken at equally-spaced times.}

Amplitude modulation of an optical lattice creates quasimomentum-selective band crossings which can be used to stitch together hybridized Floquet-Bloch bands in a variety of ways, allowing robust tunable modification of transport phenomena. We report a series of experiments probing and controlling transport of ultracold bosonic lithium atoms in Floquet-Bloch bands and demonstrating that transport in these dynamical bands can be understood in the context of a quasi-static effective band structure. Floquet hybridization in the presence of an applied force can be used to generate coherent transport over thousands of lattice sites, switch on and off Bloch oscillations, and tune the band dispersion by manipulating drive parameters. As we demonstrate, experimental measurements of dynamical evolution enable direct imaging of Floquet-Bloch band structure. 

Our experimental platform for Floquet band engineering is a degenerate quantum gas of $^7$Li in an amplitude-modulated optical lattice and an applied  harmonic magnetic potential (see Fig.~\ref{fig:fig1}f).  Each experiment begins by producing a Bose condensate of approximately $10^5$ $^7$Li atoms in the $\left | F=1,m_F=1\right\rangle$ hyperfine state in a crossed optical dipole trap centered at $x=0$. After the final stage of cooling, a Feshbach magnetic field is tuned to the shallow scattering length zero-crossing near 543.6~G~\cite{Pollack2009} to eliminate interatomic interactions, and the condensate is adiabatically loaded into a Heisenberg-limited quasimomentum distribution in the ground band of a retro-reflected optical lattice with an 85~$\mu$m beam waist.  The static lattice depth $V_0$ is $5.4\,E_R$ unless otherwise specified, where $E_R=\hbar^2k_L^2/2m$ is the recoil energy, $k_L=2\pi/\lambda$ is the lattice wavevector, $\lambda=1064$~nm is the lattice wavelength, and $m$ is the atomic mass. The curvature of the Feshbach field generates harmonic confinement with trap frequency $\omega=2\pi\times15.5$~Hz.  A field gradient from push coils translates the trap center $x_0$ such that resulting force $F(x)=-m\omega^2\left(x-x_0\right)$ drives transport in the lattice direction once the much-tighter optical dipole trap which pins the atoms to $x=0$ is removed. Dynamics are initiated by switching off the optical dipole trap and simultaneously turning on the lattice modulation using an acousto-optic modulator; this quenches the atomic ensemble into the hybridized Floquet-Bloch band. If the hybridizing quasimomentum $q^*$ is sufficiently different from the initial quasimomentum we do not observe heating from the quench, although in some cases projection of a small fraction of the atoms to higher bands can occur via multiphoton transitions. After variable hold time in the modulated lattice, the atomic position distribution is measured by {\it in-situ} absorption imaging.  

%The initial Heisenberg-limited quasimomentum distribution is centered around zero. 

Floquet-Bloch band properties and dynamics are calculated numerically by computing the eigenvalues and eigenstates of the hermitian generator of the single-period time evolution operator~\cite{Holthaus2016}. The system can be described by the 1D time-dependent Hamiltonian
\begin{equation}
\begin{split}
H=&-\frac{\hbar^2}{2m}\partial_x^2-V_0\left(1+\alpha \sin(2\pi\nu t)\right)\cos^2\left(\frac{\pi x}{d}\right)\\[1.2ex] 
&+\frac{1}{2}m\omega^2\left(x-x_0\right)^2,\label{eq:H}
\end{split}
\end{equation}
where $x$ is position along the lattice, $d$ is the lattice spacing, $m$ is the atomic mass of $^7$Li, $V_0$ is the static lattice depth, $\alpha$ is the modulation strength, and $\nu$ is the modulation frequency.   Transverse degrees of freedom play no role in the dynamics we report.  In the absence of the modulation and for weak force, the spectrum of equation~(\ref{eq:H}) consists of Bloch bands with energy $\left\{E_j(q)\right\}$, where $j$ is the band index and $q$ the quasimomentum. Amplitude modulation satisfying an interband resonance of the $i$th and $j$th band $nh\nu=\left(E_i(\pm q^*)-E_j(\pm q^*)\right)$, where $q^*$ is the resonant quasimomentum and $n$ the photon number, hybridizes the static spectrum into quasienergy bands $\{\widetilde{E}_i(q)\}$. The wavepacket center-of-mass dynamics are dictated by the local force and the group velocity of the hybridized band: $dq/dt=F$ and $dx/dt=d\widetilde{E}/dq$.  While the drive can in principle hybridize any set of bands with arbitrary order $n$, this work  focuses on the case of single-photon resonant hybridization of the ground band with the $j$th excited band, which we denote as $(0,j)$ hybridization.  %This notation, while non-unique, emphasizes the band hybridizations that govern the observed dynamics and is sufficient to specify the relevant hybrid bands that have maximal overlap with the static ground band at the center of the Brillouin zone.

\begin{figure}[t]
\centering
\includegraphics[width =\linewidth]{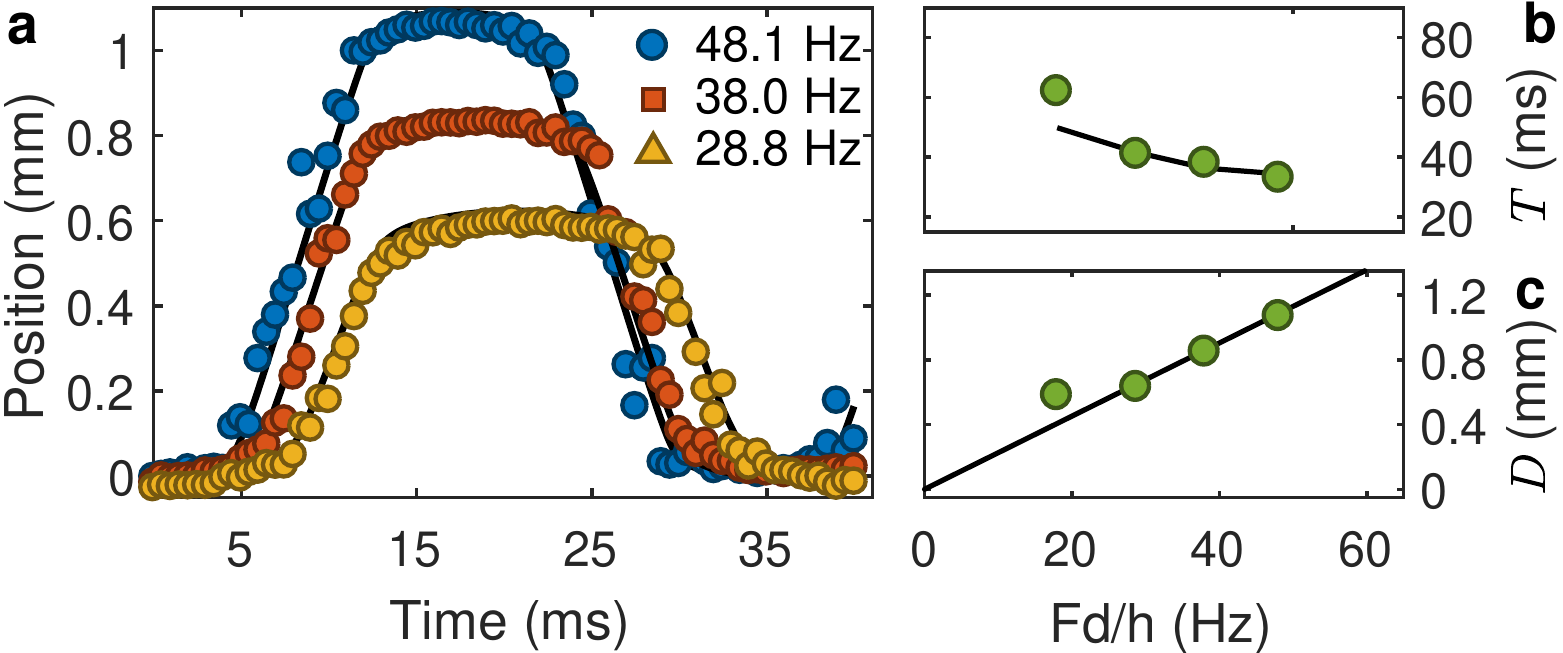}
\includegraphics[width =\linewidth]{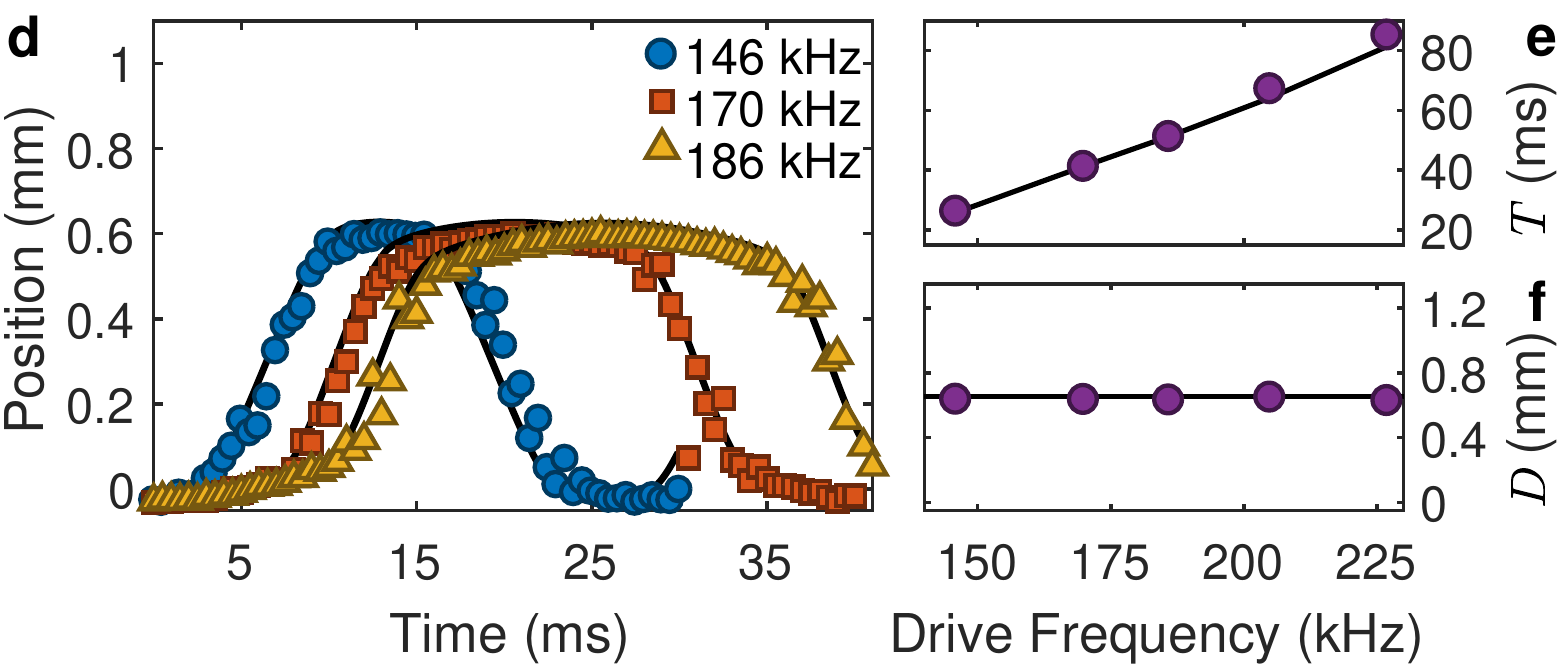}
\includegraphics[width =\linewidth]{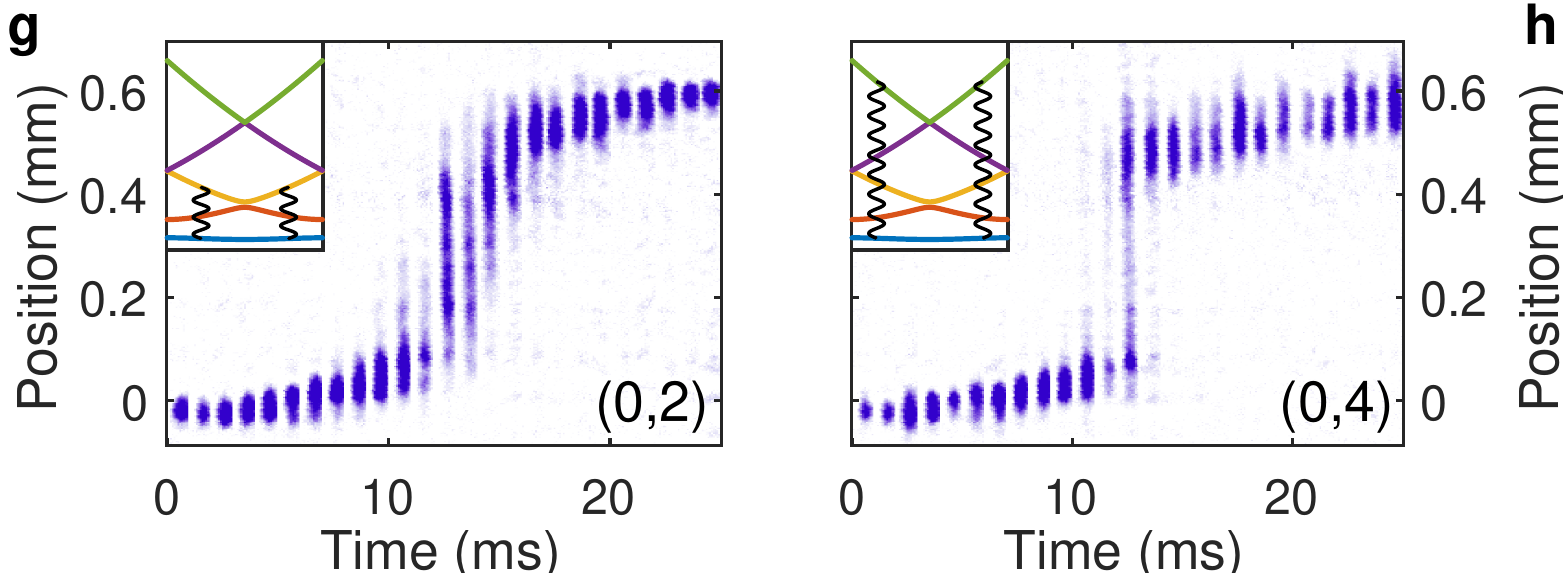}
\caption{Tunable transport in Floquet-Bloch bands. 
Panels a-g show measurements in a $(0,2)$ hybrid band with $\alpha=0.2$. 
\textbf{(a)} Atomic center-of-mass position versus time for constant drive frequency $\nu=170$~kHz and varying initial force, corresponding to the Bloch frequencies indicated in the legend. In panels a-f, theoretical expectations from the calculated Floquet-Bloch band structure are plotted as solid lines, with no fit parameters, and calculated uncertainty intervals are smaller than plotted points.  \textbf{(b,c)} Transport period and total transport distance as a function of initial force.  \textbf{(d)} Atomic position versus time for three modulation frequencies as indicated in the legend, at a constant initial force corresponding to a 26.5~Hz Bloch frequency. \textbf{(e,f)} Transport period and total transport distance as a function of drive frequency $\nu$. \textbf{(g)} Time sequence of position-space distribution in a $(0,2)$  Floquet-Bloch band with $\nu=186~$kHz and $\alpha=0.2$. 
\textbf{(h)} Time sequence of position-space distribution in a $(0,4)$ Floquet-Bloch band with $\nu=530~$kHz and $\alpha=1$. Insets in (g) and (h) indicate hybridization in the static band structure. }
\label{fig:fig2}
\end{figure}

%\textbf{(d)} Atomic position versus time for a constant initial force per atom $F=3.3\times 10^{-26}\,$N and varying drive frequency as indicated in the legend.

Fig.~\ref{fig:fig1} demonstrates the dramatic difference between transport  in the static ground band and transport in a $(0,2)$ Floquet-Bloch band.  In the absence of modulation, the local force $F$ induces Bloch oscillations (Fig.~\ref{fig:fig1}a) whose amplitude in position space is proportional to the static bandwidth~\cite{Geiger2018}. Amplitude modulation at frequency $\nu=170$~\text{kHz} and amplitude $\alpha=0.2$ hybridizes the ground and second excited band at quasimomentum $q^*\approx0.4~\hbar k_L$ as diagrammed in Fig.~\ref{fig:fig1}c.  In the Floquet-Bloch band, instead of exhibiting Wannier-Stark localization, the atomic ensemble undergoes rapid coherent oscillatory transport across approximately 2000 lattice sites (Fig.~\ref{fig:fig1}b). 

This striking transport behavior is a direct consequence of the hybridized structure of the Floquet-Bloch band. As the quasimomentum evolves, the group velocity sharply increases at the high-curvature point of the Floquet-Bloch band where $\left|q(t)\right|=q^*$. The resulting rapid transport carries the ensemble thousands of lattice sites in real space. On such large length scales, the force due to the harmonic confinement is no longer approximately constant; the ensemble moves across the entire applied potential, gaining and then losing quasimomentum without reaching the edge of the Brillouin zone.  At the position where the potential energy is the same as it was at the first point of high band curvature, energy conservation requires that the ensemble again has quasimomentum $q^*$, and the group velocity sharply decreases. The microscopic dynamics at the high-curvature point involve exchange of photons with the lattice, as occurs for example in pulsed Bragg acceleration techniques used in atom interferometry to increase free-space momentum~\cite{BObeamsplitterforAI}. Both direct measurement and Landau-Zener calculations indicate that deviations from perfect fidelity of transfer between static bands at the Floquet-induced avoided crossings are below one part in $10^{4}$ for all data we report. Transport in the hybridized band is characterized by periods of rapid transfer across the entire trapping region connected by relatively slow Bloch-oscillation-like motion at the turning points. Variations in ensemble width during transport can arise from both force inhomogeneity~\cite{Geiger2018} and from the modified dispersion~\cite{rho}. The full position-velocity evolution is shown in Fig.~\ref{fig:fig1}e. Notable features of the observed dynamics include high-fidelity long-range transport of nearly all atoms in the condensate, coherent Bloch oscillation dynamics at opposite ends of the trap, maximum transport velocities far in excess of those characterizing the bare harmonic potential, and a high degree of control attainable by varying drive properties.

To further probe the dynamics we study the dependence of long-range transport on initial applied force, drive frequency, and hybridized band indices.  As shown in Fig.~\ref{fig:fig2}a and b, the total transport distance $D$ increases with increasing force while the oscillation period $T$ decreases.  The observed behavior agrees quantitatively with fit-parameter-free numerical calculations for the Floquet-Bloch band shown as solid black lines in Figs.~\ref{fig:fig2}a, b, and c. Larger total transport distances can be achieved by reducing the harmonic trap frequency or increasing the applied force, although the latter method will eventually be limited by Bragg scattering.  Figs.~\ref{fig:fig2}d, e, and f show the results of varying the hybridizing frequency $\nu$ at constant force.  Increasing $\nu$ hybridizes the bands closer to the edge of the Brillouin zone, which increases the time to reach $q^*$ so that the oscillation period increases while the transport distance remains fixed.  Again, the observed wavepacket evolution agrees quantitatively with fit-parameter-free numerical predictions, demonstrating the effectiveness of the Floquet-Bloch formalism for describing this controllable long-distance transport. 

\begin{figure}[t]
\centering
\includegraphics[width=\linewidth]{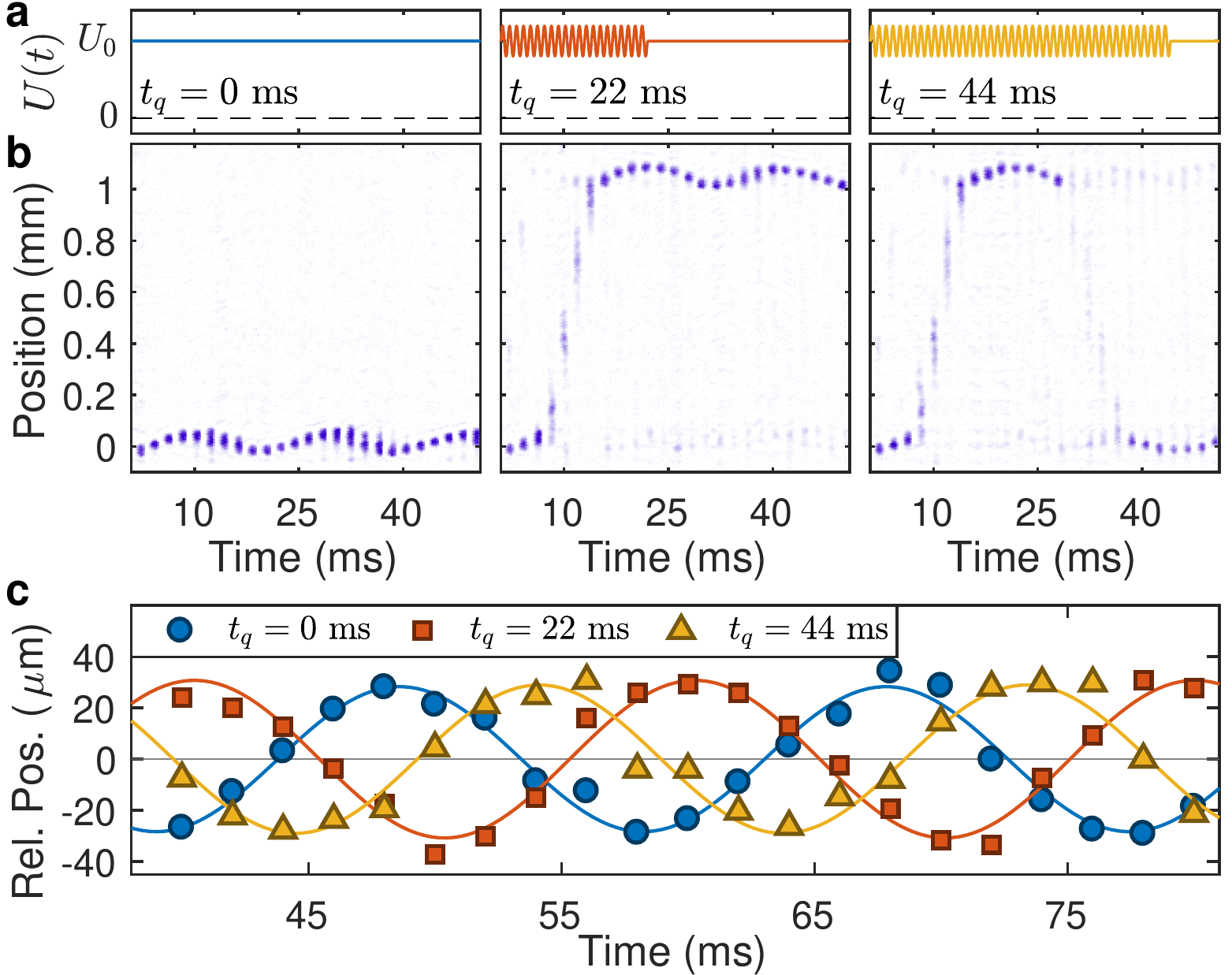}
\caption{Transport control via Floquet quenches with $\nu=170$~kHz and $\alpha=0.4$.  {\bf (a)} Quench protocol for modulation times of 0, 22, and 44~ms.  {\bf (b)} Time sequence of images of atomic ensemble for the three quench protocols.   {\bf (c)} Relative position evolution of the atomic ensemble after all quenches.  Solid lines are sinusoidal fits, and estimated uncertainty in measured data is smaller than plotted points.  }
\label{fig:toggle}
\end{figure}

Floquet hybridization of different pairs of static bands gives rise to distinct transport properties. Fig.~\ref{fig:fig2}g and \ref{fig:fig2}h compare transport dynamics in $(0,2)$ and $(0,4)$ hybrid bands. Evolution in the $(0,4)$ band leads to dramatically faster long-range transport, due to the increased group velocity and band curvature.  During this evolution, the ensemble stretches across the entire extent of the trapping potential, but still returns to the static ground band at the far edge of the trap. 

Diabatic quenches between static bands and hybridized Floquet bands provide a powerful experimental tool for dynamical control of transport properties. Quenching back to the static lattice after a total modulation time $t_q$ projects the Floquet-Bloch state back onto the original static spectrum.  Fig.~\ref{fig:toggle} shows the results of such quenched modulation experiments in which the modulation depth $\alpha$ is set suddenly to zero after some variable evolution time in the Floquet-Bloch band.  Quenching near the turning points of the position-space oscillation projects the atomic ensemble back onto the Wannier-Stark localized ground band, at a position which can vary by thousands of lattice sites depending on $t_q$.  Quenching into an excited static band is also possible; a recent study demonstrated that a related technique can be used to experimentally realize a relativistic harmonic oscillator~\cite{rho}.
The small fraction of atomic population away from the main ensemble visible in Fig.~\ref{fig:toggle}b arises from multiphoton projection to higher bands during the initial quench.
The identical amplitude and frequency of position-space Bloch oscillations after the three distinct quench protocols in Fig.~\ref{fig:toggle}c indicates the non-dissipative nature of quenched Floquet-Bloch transport.  
%The relative phase shifts of the position-space Bloch oscillations are consistent with expectations based on the time spent in the hybridized state and the changing sign of the force during the transport. 

\begin{figure}[t]
\centering
\includegraphics[width =\linewidth]{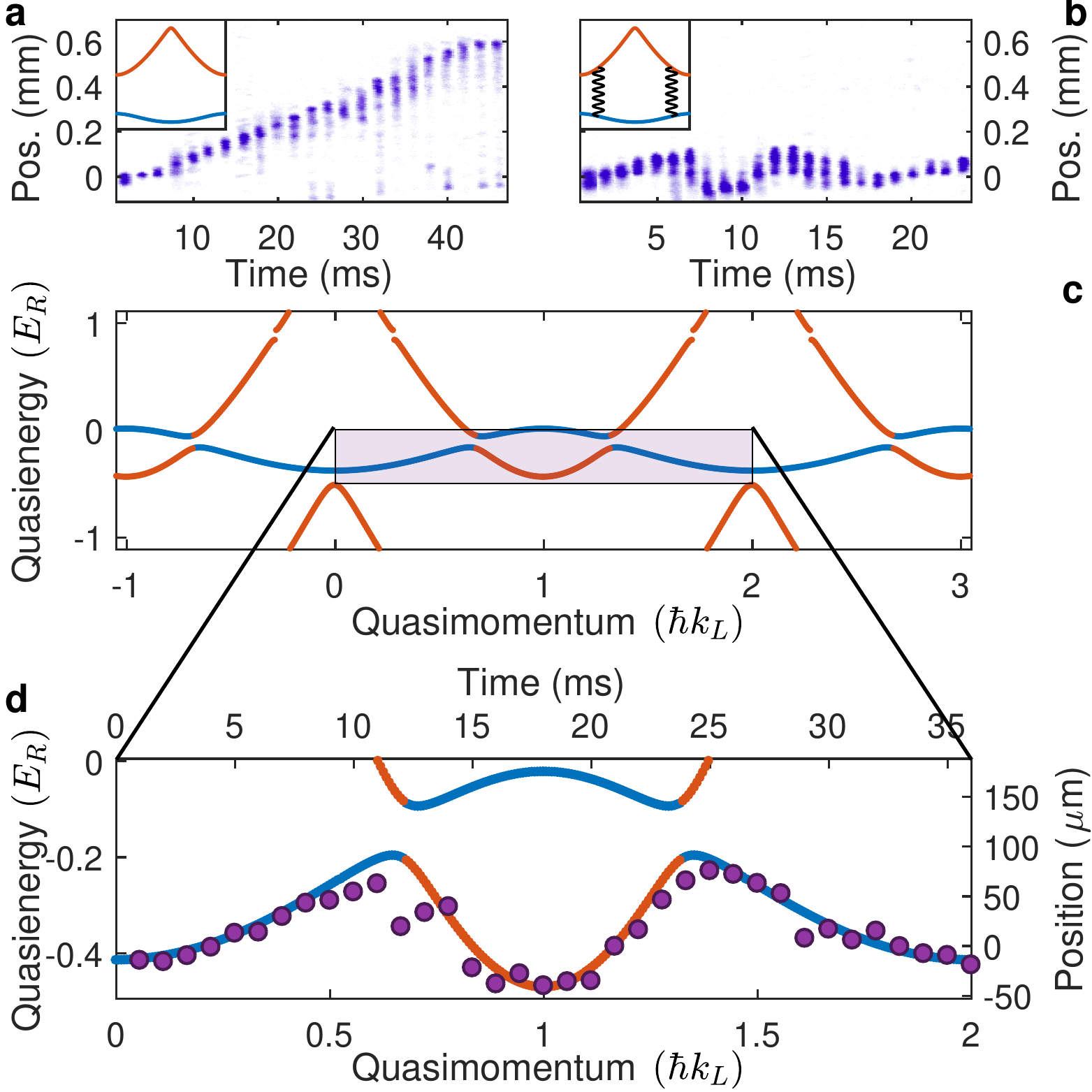}
\caption{Imaging a Floquet-Bloch band.  {{\bf (a)}} Time sequence of images of atoms in a static band with reduced lattice depth $V_0=3.6~E_R$, subjected to an initial force per atom corresponding to a Bloch oscillation frequency of 28.9~Hz.  Inset is the calculated band structure. {{\bf (b)}} Evolution in a $(0,1)$ hybrid Floquet-Bloch band with $\nu=56~$kHz, $\alpha=0.25$, and the same lattice depth as (a).  Inset is the static band structure with rippled lines indicating the resonant coupling.
{{\bf (c)}} Calculated quasienergy spectrum in the extended zone scheme for the $(0,1)$ hybrid band of (b). Color corresponds to the static band with maximal probability overlap with the Floquet state according to the band colors in (a). Shaded region corresponds to the mapped part of the Floquet band in (d).
{{\bf (d)}} Comparison of the real-space evolution in (b) to the Floquet spectrum in (c) according to the mapping of equation~(\ref{eq:dE}) with no fit parameters.  Measurement uncertainty is smaller than plotted points.  The atomic motion images the Floquet-Bloch band dispersion.}
\label{fig:p}
\end{figure}

Hybridizing bands of opposite parity gives rise to qualitatively different phenomena. While opposite-parity coupling is forbidden at $q^*=0$ due to the even-parity nature of amplitude modulation, hybridization at finite quasimomentum is both allowed and observed. Fig.~\ref{fig:p} shows the result of $(0,1)$ coupling at $q^*\approx 0.66~\hbar k_L$, using amplitude modulation with $\nu=56~$kHz, $\alpha=0.25$, and $V_0=3.6~E_R$.  At this reduced lattice depth, atoms in the unmodified ground band do not Bloch oscillate but undergo ballistic transport in the trapping potential, never reaching the Brillouin zone edge~(Fig.~\ref{fig:p}a). Quenching into the $(0,1)$ hybrid band causes Wannier-Stark localization due to Bloch oscillations in the Floquet-hybridized band~(Fig.~\ref{fig:p}b).  Here the opposite curvature of the two static bands results in a flatter hybrid band with smaller bandwidth and local extrema in the dispersion, as shown in Fig.~\ref{fig:p}c.  Deviations from unit Landau-Zener fidelity for Bloch oscillations near the avoided crossings are calculated to be below a part per million for this drive amplitude and Bloch frequency, and are not  observed.

Strikingly, these oscillations enable direct imaging of the Floquet-Bloch band structure. A recent experiment~\cite{Geiger2018} demonstrated that for a constant force, the position evolution of an atomic ensemble undergoing Bloch oscillations in a {\em static} band constitutes a direct image of the energy-momentum dispersion relation according to the mapping:
\begin{align}
E&=F x,\hspace{.3in}  q=F t. \label{eq:dE}
\end{align}
This can be intuitively understood by considering that the group velocity is equal to both $dE/dq$ and $dx/dt$. Fig.~\ref{fig:p}d experimentally demonstrates that this mapping can be generalized to Floquet-Bloch bands by comparing the center-of-mass motion in a $(0,1)$ hybrid band to the calculated band dispersion of Fig.~\ref{fig:p}c, scaled according to equation~(\ref{eq:dE}). There are no fit parameters in this plot, as the force is measured independently.  The close agreement between the measured atomic position and the band dispersion demonstrates direct imaging of a hybridized Floquet-Bloch band.   
%The small Bloch oscillation frequency of 28.9~Hz relative to the recoil frequency $E_R/h=25.18$~kHz strongly suppresses Landau-Zener transitions to the complementary hybridized band.  DISCUSS LZ DIFFERENTLY

% {{\bf (a)}} Time sequence of images of atoms undergoing ballistic transport in a static band with reduced lattice depth $V_0=3.6~E_R$, subjected to an initial force per atom of 3.6$\times 10^{-26}$~N. Inset is the calculated band structure.

This Floquet-Bloch band image is reminiscent of time- and angle-resolved photoemission maps of Floquet bands in laser-driven topological insulators~\cite{gedki-floquetbloch}. This analogy suggests  potentially fruitful connections between the results we present and topics of current interest in condensed matter, including the effect of Bloch oscillations and inter-band transitions on high-harmonic generation in crystals and prospects for all-optical band structure reconstruction in solids~\cite{Ghimire-HHGsolids,
sherwin-recollision,ivanov-HHGbandstructure,huber-HHGblochoscs,Huber-realtimehhginsolids,reis-solidHHG,
corkum-opticalbandreconstruction}. Using techniques like those we present, cold atom quantum simulation experiments could serve as a complementary tool for exploration of band dynamics, probing and realizing phenomena at the edge of current ultrafast experimental capabilities.  

The controlled Floquet-Bloch transport dynamics presented here also open the path to the realization and study of more complex phenomena, including polychromatic driving for hybridization of multiple bands at multiple quasimomenta, Floquet-based creation of topologically nontrivial bands~\cite{refaelgalitski-floquettopologicalinsuator,torres-floquetTI}, and the controlled introduction of disorder. The enhanced control of band structure and transport demonstrated here may also be useful for lattice-based metrology. One possibility along these lines would be continuously-trapped atom interferometry using wavepackets split by a large distance in a $(0,2n)$ hybrid band for some $n$.  While this study has focused on coherent dynamics in the absence of interaction-induced dephasing, the introduction of tunable interatomic interactions would open up a broader array of possibilities in many-body Floquet engineering; some of these are explored in recent work using the same apparatus~\cite{floquetarxiv}.

In summary, we have demonstrated tunable coherent control of long-range quantum transport in hybridized Floquet-Bloch bands.  We have used hybridization of various pairs of four separate static bands at varying quasimomenta to realize rapid long-distance transport of a Bose condensate, switchable Wannier-Stark localization, and direct imaging of a hybrid Floquet-Bloch band. 

\section{Acknowledgments}
We thank Toshihiko Shimasaki, Peter Dotti, Sean Frazier, Ethan Simmons, James Chow, Shuo Ma, and Yi Zeng for experimental assistance, Gil Refael for useful discussion, and Mark Sherwin for a critical reading of the manuscript, and acknowledge support from the Army Research Office (PECASE W911NF1410154 and MURI W911NF1710323), National Science Foundation (CAREER 1555313), and the UC Office of the President (CA-15-327861).

\bibliography{scibib,Mendeley,dmwbibliography,0kurtbib}
\bibliographystyle{apsrev4-1}

\end{document}